\documentclass[letterpaper,print,prb,numerical,showpacs, showkeys]{revtex4-1}

\usepackage{graphicx}

\begin{document}

\title{The Thermodynamics and the Inverse Isotope Effect of superconducting PdH and PdD under pressure}

\author{S. Villa-Cort\'es and R. Baquero}

\affiliation{Physics Department, Cinvestav-IPN\\
 Av. IPN 2508 GAM, 07360 Ciudad de M\'exico, M\'exico}

\date{\today}

\begin{abstract}

We present in this paper the thermodynamics of superconducting PdH and PdD under pressure. We make use of a general method to calculate the thermodynamics under pressure within the Migdal-Eliashberg theory. We have considered the crystal lattice to be zincblende taking into account the experimental evidence for both PdH and PdD at temperatures below 55 K. We have studied, in particular, the changes induced by pressure in the critical temperature, $T_c$, in the specific heat jump at $T_c$, in the energy gap at $T=0K$, in the deviation function $D(t)$ and in the isotope effect coefficient, $\alpha$. We get a very good agreement with experiment where this data exist. This method represents a basis on which the thermodynamics of other hydrides under pressure can be calculated.

\end{abstract}




\maketitle


\section{INTRODUCTION}

In general, the volume of a solid is reduced when hydrostatic pressure is applied and as a result of the general stiffening of the lattice, the phonon spectrum shifts upwards to higher frequencies. This fact has an important influence on the properties of the solid. In particular, in superconductors, pressure induces changes in the whole thermodynamics. The critical temperature, $T_c$, a very important parameter to decide on applications, can either increase or decrease. There is no criterion at present that would allow to predict the actual behaviour from the knowledge of the characteristics of the system at ambient pressure. For example, $T_c$ decreases under  pressure for PdH, Pb, Sn and Hg \cite{TROFIMENKOFF1969661,JClark,PhysRevB.39.4110} but increases in an important way in sulfur hydride, $H_{3}S$. This system holds the record at present as a high-Tc superconductor ($T_c \approx 200K$) \cite{203}. 

Another important parameter is the isotope effect coefficient, it has been instrumental in understanding the mechanism responsible for Cooper pair formation in conventional superconductors. There is no general behavior for the isotope effect coefficient. In $MgB_{2}$ \cite{MgB2} and the fullerides \cite{PhysRevLett.83.404}, for example,  the isotope coefficient is substantially reduced from the BCS value. In PdH, the compound of interest in this work, the isotope coefficient is large and negative $(\alpha\approx-0.3)$ \cite{CHEN1989485} and under pressure it diminishes steadily \cite{PhysRevB.39.4110}. This is in contrast to $H_{3}S$  between 130 GPa and 140 GPa where the isotope coefficient goes down very quickly and then it goes further down but more slowly \cite{0953-2048-30-4-045011,italiano}.Further, $^{6}Li$  exhibits an unusually large isotope effect below 21 GPa and between 21 to 26 GPa, the superconducting isotope effect becomes inverse\cite{Schaeffer06012015}. As a function of the doping concentration, the Oxygen and boron isotope coefficients in $La_{2-x}Sr_{x}CuO_{4}$ \cite{Bianconi1,Bianconi2,Bianconi3} and in $Mg_{1-x}Al_{x}B$ \cite{Bianconi7} respectively also show a variation from the BCS value.

In this paper, we will develop a method to calculate the changes under pressure in some important thermodynamic functions as the specific heat jump at $T_c$, $\Delta C_v(T_c)$, the energy gap at $T=0$, $\Delta_0$, the thermodynamic deviation function $D(t)$, as well as the isotope effect coefficient, $\alpha$, for PdH(D). This method uses the solution of the linearized Migdal-Eliashberg equations and the functional derivative of $T_c$ with the Eliashberg function.

In the past decade, numerous theoretical predictions have been made on the structure and superconducting behavior of stoichiometric and hydrogen-enriched hydrides with a variety of elements at high pressures. The surprisingly high observed $T_c$ in $H_{3}S$ raises the possibility that even higher transition temperatures may be attainable in hydrides. It is desirable to establish general characteristics that allow the understanding of the underlying mechanisms of high-$T_c$ superconductivity in hydrogen-rich materials \cite{PhysRevB.96.100502}. To calculate in detail the behavior of both PdH and PdD under pressure constitutes a basis to identify these characteristics.

The rest of the paper is organized as follows. In the next section II, we present the theory and the basic equations that sustain our work. In section III we present the technical details. Section IV is devoted to our results and in a final section V, we draw our conclusions.

\section{THEORY AND BASIC EQUATIONS}

In this work, we use the solution of the linearized Migdal-Eliashberg equation (LMEE) \citep{Daams1979,PhysRevB.12.905} valid at $T_{c}$. For an isotropic superconductor, the LMEE is

\begin{eqnarray}
\rho\bar{\Delta}_{n} & = & \pi T\sum_{m}\left[\lambda_{nm}-\mu^{*}-\delta_{nm}\frac{\left|\tilde{\omega}_{n}\right|}{\pi T}\right]\bar{\Delta}_{m}.
\label{Eli Ecua}
\end{eqnarray}

Where $\bar{\Delta}_{n}$ is given by
 
\begin{eqnarray}
\bar{\Delta}_{n} & = & \frac{\left|\tilde{\omega}_{n}/\omega_{n}\right|\Delta_{n}}{\left|\tilde{\omega}_{n}\right|+\pi T\rho},
\end{eqnarray}

Here $\rho$ is the breaking parameter that becomes zero at $T_{c}$. The
frequency $\tilde{\omega}_{n}$ is 

\begin{eqnarray}
\tilde{\omega}_{n} & = & \omega_{n}+\pi T\sum_{m}\lambda_{nm}sig(\omega_{m}),
\end{eqnarray}

and $i\omega_{n}$ are the Matsubara frequencies defined on the imaginary axis, $i\omega_{n}=i\pi T\left(2n+1\right)$ with $n=0,\pm1,\pm2\ldots$. The e-ph coupling parameter $\lambda_{nm}$ is defined as 

\begin{eqnarray}
\lambda_{nm} & = & 2\int_{0}^{\infty}\frac{d\omega\omega\alpha^{2}F\left(\omega\right)}{\omega^{2}+\left(\omega_{n}-\omega_{m}\right)^{2}}.\label{eq:lambdas}
\end{eqnarray}

 $\lambda_{nn}$ is the known electron-phonon interaction parameter and $\mu^{*}$ is the electron-electron repulsion parameter.
The Eliashberg function is defined as

\begin{eqnarray}
\alpha^{2}F\left(\omega\right)= &  & \frac{1}{N\left(\epsilon_{F}\right)}\sum_{nm}\sum_{\vec{q}\nu}\delta\left(\omega-\omega_{\vec{q}\nu}\right)\sum_{\vec{k}}\left|g_{\vec{k}+\vec{q},\vec{k}}^{\vec{q}\nu,nm}\right|^{2}\label{eq:a2f-def}\\
 &  & \times\delta\left(\epsilon_{\vec{k}+\vec{q},m}-\epsilon_{F}\right)\delta\left(\epsilon_{\vec{k},n}-\epsilon_{F}\right),\nonumber 
\end{eqnarray}

where $g_{\vec{k}+\vec{q},\vec{k}}^{\vec{q}\nu,nm}$ are the matrix elements of the electron-phonon interaction, $\epsilon_{\vec{k}+\vec{q},m}$ and $\epsilon_{\vec{k},n}$ are the energy of the quasi-particles in bands $m$ and $n$ with vectors $\vec{k}+\vec{q}$ and $\vec{k}$ respectively. The functional derivative of the critical temperature with respect to the Eliashberg function is given as \cite{Bergmann1973}

\begin{eqnarray}
\frac{\delta T_{c}}{\delta\alpha^{2}F\left(\omega\right)} & = & -\left(\frac{\partial\rho}{\partial T}\right)_{T_{c}}^{-1}\frac{\delta\rho}{\delta\alpha^{2}F\left(\omega\right)}.\label{Dev F}
\end{eqnarray}

To study the thermodynamics at temperatures below the critical temperature at each pressure we use the nonlinear Migdal-Eliashberg equation (NLMEE)

\begin{eqnarray}
\frac{\left|\tilde{\omega}_{n}\right|}{\pi T}\bar{\Delta}_{n} & = & \sum_{m}\left(\lambda_{nm}-\mu^{*}\right)\frac{\bar{\Delta}_{m}}{\left(1+\bar{\Delta}^2_{m}\right)^{1/2}},\label{NEli}
\end{eqnarray}

\begin{eqnarray}
\tilde{\omega}_{n} & = & \omega_{n}+\pi T\sum_{m}\frac{\lambda_{nm}}{\left(1+\bar{\Delta}^2_{m}\right)^{1/2}}sig(\omega_{m}).
\end{eqnarray}

From the solution of Eq.(\ref{NEli}) the free energy difference $\Delta F(T)$ between the superconducting and normal states can be computed. It is given by

{\footnotesize{}
\begin{eqnarray}
\Delta F(T)= &  & N(0)\pi T\sum_{n}\left[2\left|\tilde{\omega}_{n}\right|\left(\frac{1}{\sqrt{1+\bar{\Delta}_{n}^{2}}}-1\right)\right.\nonumber \\
 &  & +\pi T\sum_{m}\left(\lambda_{nm}-\mu^{*}\right)\frac{\bar{\Delta}_{n}}{\sqrt{1+\bar{\Delta}_{n}^{2}}}\frac{\bar{\Delta}_{m}}{\sqrt{1+\bar{\Delta}_{m}^{2}}}\\
 &  & \left.-\pi T\sum_{m}\left(1-\frac{1}{\sqrt{1+\bar{\Delta}_{n}^{2}}}\frac{1}{\sqrt{1+\bar{\Delta}_{m}^{2}}}\right)\lambda_{nm}sgn\left(\omega_{m}\omega_{n}\right)\right]\nonumber 
\end{eqnarray}
}{\footnotesize \par}

 $N(0)$ is the single-spin density of electronic states at the Fermi energy. The other quantities we are interested in below $T_c$ are the critical magnetic field $H_c(T)$ which is given by 

\begin{eqnarray}
H_c(T)=\left(8\pi \left|\Delta F \right| \right)^{1/2},
\end{eqnarray}

the critical magnetic field deviation function, $D(t)$, given by

\begin{eqnarray}
D(t) = H_c(T)/H_c(0) - (1-t^2)
\end{eqnarray}
where $t=T/T_c$. The difference between the superconducting (S) and normal state (N) specific heat at constant volume, $\Delta C_v$, follows from the second derivative of the free energy

\begin{eqnarray}
\Delta C_v(T) = -T(d^{2}\Delta F/dT^{2})
\end{eqnarray}

\section{TECHNICAL DETAILS}

Quite an amount of theoretical and experimental work has been done since the discovery of superconductivity in PdH and PdD. Most of the work done so far considers the rocksalt crystalline structure where the hydrogen atoms are located on the octahedral sites of the fcc lattice of Palladium \cite{PhysRevB.14.3630,YUSSOUFF1995549,CHEN1989485,BROWN197599,CRESPI1992427,PhysRevLett.57.2955,PhysRevLett.35.110,PhysRevB.17.141,0022-3719-7-15-015,PhysRevB.39.4110,Kara,PhysRevB.45.12405,PhysRevB.29.4140,PhysRevLett.34.144,PhysRevB.12.117,PhysRevLett.111.177002}. However, there is another possibility. Neutron diffraction techniques have been employed to study the hydrogen-atom configuration in a single-phase sample of beta-PdH at several selected temperatures. The suggested low-temperature ($T\ll55$ K) structure of this compound is one which conforms to the space group $R\bar{3}m$. This means that, depending on temperature ($T\ll55$ K), the hydrogen atoms move from their octahedral positions towards tetrahedral ones forming the zincblende structure  \cite{PhysRev.137.A483,CAPUTO-ALI,PhysRevB.78.014104,doi:10.1063/1.4901004}. For PdD something very similar occurs \cite{0295-5075-64-3-344}.  In a theoretical study, for pressures below 20 GPa at 0 K  the stable structure was found to be zincblende \cite{doi:10.1021/jp210780m}. So, following the facts just mentioned, in this paper we will consider PdH and PdD in the zincblende crystal structure that has not been yet considered.

\subsection{Computational Details}

The phonon spectra and the Eliashberg function where calculated within the density functional perturbation theory \cite{0953-8984-21-39-395502,RevModPhys.73.515} as implemented in the Quantum - Espresso suite code \cite{0953-8984-21-39-395502}. We use the scalar relativistic pseudo potentials of Pardue and Zunger (LDA) \cite{PhysRevB.23.5048}, a 150 Ry cutoff for the plane-wave basis and a 32 X 32 X 32 mesh for the BZ integration in the unit cell. For the force constants matrix we used a 16 X 16 X 16 mesh. The sum over $\vec{k}$ in Eq. (\ref{eq:a2f-def}) required a 72 X 72 X 72 grid.

To solve the LMEE, we used a cut-off frequency, $\omega_{cutoff} = 10 \omega_{ph}$ where $\omega_{ph}$  is the maximum phonon frequency, to cut the sum over the Matsubara frequencies. 

\subsection{The phonon spectra and the Eliashberg functions}

\begin{figure}[h]
\begin{center}
\includegraphics[width=8.4cm]{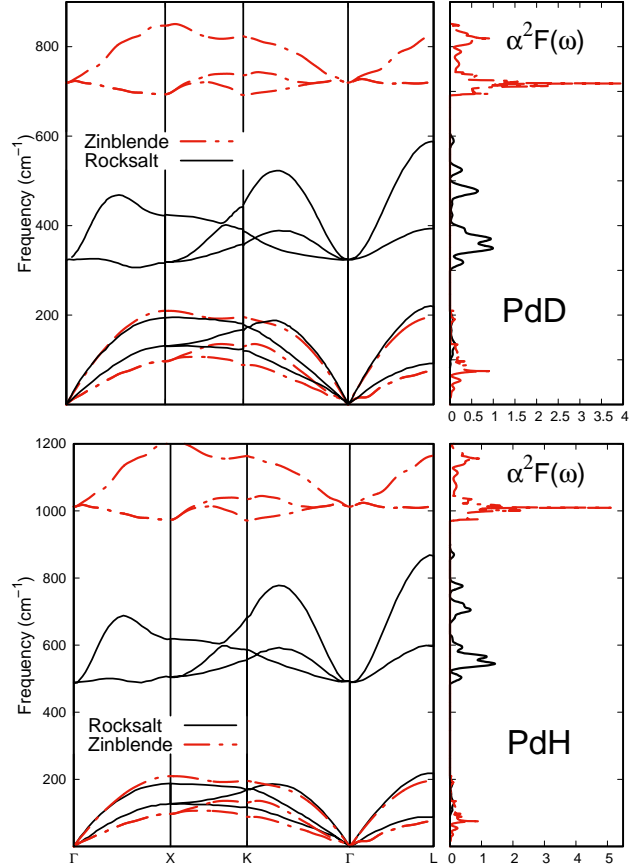}
\caption{\label{fig1}(color online) Phonon spectra and Eliashberg function calculated for PdH and PdD with the zincblende crystal structure. We also include the ones for the rocksalt crystal structure taken from reference \cite{PhysRevLett.111.177002}.}
\end{center}
\end{figure}

In figure Fig. \ref{fig1} we show our calculated phonon spectrum and the Eliashberg function for the zincblende crystal structure. The phonon spectrum and the Eliashberg function for the rocksalt structure are also shown, the data were taken from reference \cite{PhysRevLett.111.177002}. For both crystal structures the phonon spectrum and therefore, the Eliashberg function, present two distinct frequency regions. The acoustic region, from the Pd vibrations, is in general similar for both structures and it ranges from 0 to approximately 250 $cm^{-1}$. In the two structures the optical frequency region from the Hydrogen and Deuterium vibrations, is separated by a frequency gap from the acoustic ones. The hydrogen optical region is higher in frequency than that of the deuterium one in both structures.  Further, the optical region in the zincblende structure is shifted to higher frequencies as compared to the rocksalt structure.

\section{THE THERMODYNAMICS}

The thermodynamics of the PdH and PdD systems under pressure has not been yet addressed theoretically. As we mention above, several theoretical works have been done for these systems at zero pressure considering the structure to be rocksalt. Here we will consider the PdH and PdD in the zincblende crystal structure. In order to check the validity of this hypothesis, we investigate the pressure dependence of the critical temperature and of the thermodynamic properties of PdH and PdD.

The method that we use to calculate $T_c$ as a function of pressure uses the functional derivative of $T_c$ with the Eliashberg function as a function of pressure and requires only to know the critical temperature at an initial pressure and the knowledge of the Eliashberg function at each pressure of interest \cite{ivan}. Once we get $T_c$ and $\mu^{*}$ we can solve the NLMEE to calculate the thermodynamics at each pressure. 

\subsection{The superconducting critical temperature}

\begin{center}
\begin{figure}
\includegraphics[width=8.4cm]{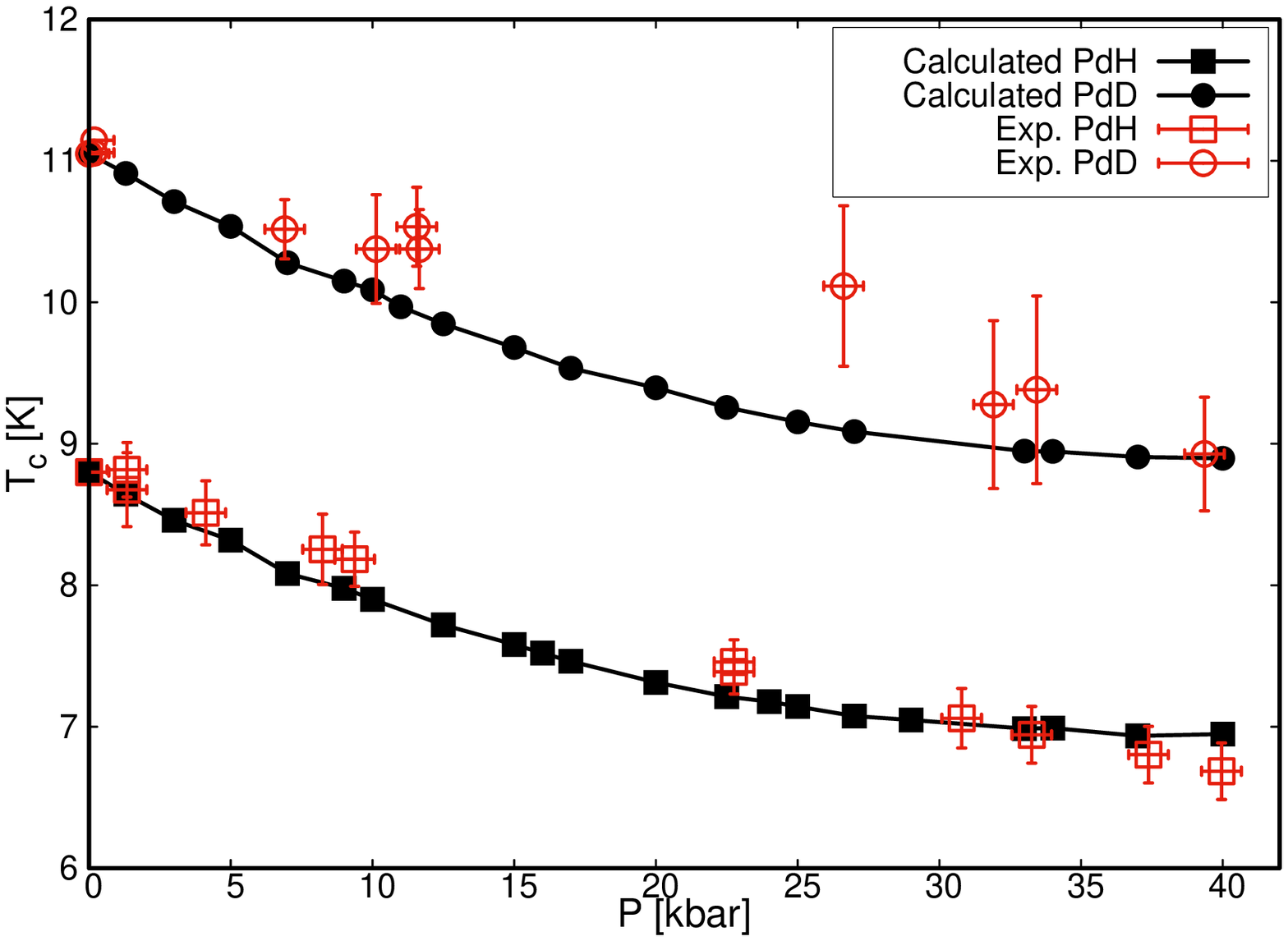}
\caption{(color online) $T_{c}$ under pressure calculated for PdH and PdD. Experimental data for PdH and PdD are also shown \citep{PhysRevB.39.4110}.}
\label{fig:Tc-P}
\end{figure}
\end{center}

To calculate $T_c$ we start at an initial pressure where the critical temperature is known, $T_{c}(P_{i})$. From it we obtain $\mu^{*}(P_{i})$ by fitting it to $T_{c}(P_{i})$ solving the LMEE. Then we get the functional derivative of $T_c$ with the Eliashberg function, $\alpha^{2}F\left(\omega,P_{i}\right)$, $\frac{\delta T_{c}}{\delta\alpha^{2}F\left(\omega\right)}$ using the formalism of Bergmann \cite{Bergmann1973} and of Leavens \cite{LEAVENS19741329}. At another pressure, say $P_{i+1}$, we calculate $T_{c}(P_{i+1})$ from \cite{ivan}
\begin{eqnarray}
T_{c}(P_{i+1})=T_{c}(P_{i})+\Delta T_{c}(P_{i})
\end{eqnarray} 
where $\Delta T_{c}(P_i)$ is given by
\begin{eqnarray}
\Delta T_{c}(P_i) & = & \int_{0}^{\infty}\frac{\delta T_{c}(P_i)}{\delta\alpha^{2}F\left(\omega\right)}\Delta\alpha^{2}F\left(\omega\right)d\omega,
\end{eqnarray}
and $\Delta\alpha^{2}F\left(\omega\right)=\alpha^{2}F\left(\omega , P_{i+1}\right)-\alpha^{2}F\left(\omega,P_{i}\right)$.

Since we already know $T_{c}(P_{i+1})$ and $\alpha^{2}F\left(\omega,P_{i+1}\right)$, we can solve the LMEE to get $\mu^{*}(P_{i+1})$. The procedure can be repeated to get $T_{c}$ at each pressure of interest. Notice that we get $T_{c}$ and $\mu^{*}$ as a function of pressure in a consistent way with Eliashberg theory. We do not use any approximation.

In Fig. \ref{fig:Tc-P}, we show the critical temperature at several pressures for PdH and PdD calculated with the method described above. We use for PdH, $T_{c}=8.8\:K$ at $0$ $kbar$ \citep{PhysRevB.39.4110} as our starting data. As it can be seen from Fig. \ref{fig:Tc-P} in the whole pressure interval that we have calculated the agreement with the experimental values is excellent.
In the case of PdD, we took as our starting temperature $T_{c}=11.05\:K$ at 0 $kbar$ \citep{PhysRevB.39.4110}. In this case, the agreement with experiment for $T_{c}$ as a function of pressure in the calculated interval follows closely the same trend as the experimental values and $\textit{grosso modo}$ agrees with it.

\begin{table}[h]
\caption{\label{tab:table1}Calculated values of the Coulomb parameter, $\mu^{*}$, and the maximum of the functional derivative of $\delta T_{c}/\delta\alpha^{2}F\left(\omega,P\right)$, $\omega_{opt}$, for PdH and PdD at several pressures.}
\footnotesize
\begin{ruledtabular}
\begin{tabular}{ccccc}
& \multicolumn{2}{c}{PdD} & \multicolumn{2}{c}{PdH} \tabularnewline
\hline 
$P\:[kbar]$ & $\mu^{*}$ & $\omega_{opt}$ [meV]& $\mu^{*}$ & $\omega_{opt}$ [meV]\tabularnewline
\hline 
0 &  0.021813 & 8.0509 & 0.061998 & 6.4245\tabularnewline
1.3 & 0.022078 & 7.9471 & 0.062238 & 6.3160\tabularnewline
3 & 0.022902 & 7.8049 & 0.063083 & 6.1785\tabularnewline
5 & 0.023242 & 7.6805 & 0.063373 & 6.0799\tabularnewline
7.5 & 0.023937 & 7.4937 & 0.064250 & 5.9120\tabularnewline
9 & 0.024261 & 7.4008 & 0.064580 & 5.8320\tabularnewline
10 & 0.024434 & 7.3548 & 0.064825 & 5.7804\tabularnewline
12.5 & 0.025109 & 7.1824 & 0.065473 & 5.6488\tabularnewline
22.5 & 0.027599 & 6.7497 & 0.068341 & 5.2752\tabularnewline
27 & 0.028827 & 6.6257 & 0.069721 & 5.1780\tabularnewline
33 & 0.030762 & 6.5197 & 0.071853 & 5.1089\tabularnewline
34 & 0.031069 & 6.5184 & 0.072217 & 5.1125\tabularnewline
37 & 0.032192 & 6.4843 & 0.073817 & 5.0639\tabularnewline
40 & 0.033314 & 6.4764 & 0.074878 & 5.0729\tabularnewline
\end{tabular}\end{ruledtabular}

\end{table}

Table \ref{tab:table1} shows the electron-electron repulsion parameter  $\mu^{*}$ which depends indirectly on the mass of the ions \cite{LEAVENS19741329}. It is reduced considerably in PdD as compared to PdH.  This means that the coulomb electron-electron repulsion weakens in PdD.

\subsection{The deviation function}

\begin{figure}[h]
\includegraphics[width=8.4cm]{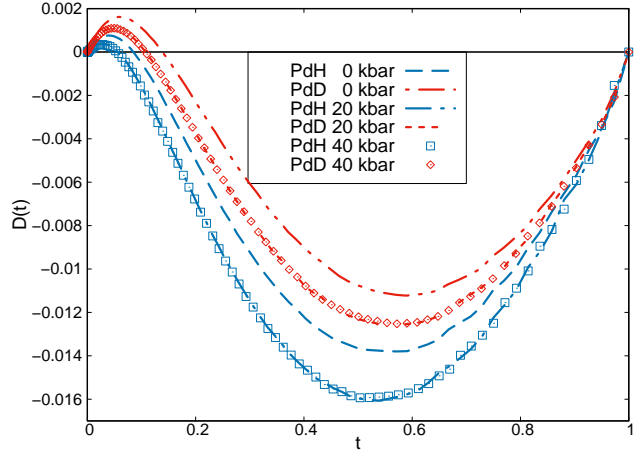}
\caption{\label{fig:Dv}(color online) Deviation function $D(t)$.}
\end{figure}

To calculate the thermodynamics from the formulas given above, we use the $\alpha^{2}F\left(\omega,P\right)$, $\mu^{*}$ and $T_{c}$ calculated at each pressure in the previous section to solve the nonlinearized Eliashberg equations (Eq. \ref{NEli Ecua}).

Figure \ref{fig:Dv} shows plots of the calculated deviation function, $D(t)$, for $0$, $20$ and $40$ Kbar pressures. This function is useful to magnify the difference between weak and strong coupling systems. In a strong coupling system, such as Pb \cite{Vashishta1975}, the deviation function is everywhere positive. We find for both systems, PdH and PdD, to be intermediate coupling and show a tendency to weaker coupling with increasing pressure between $0$ and $20$ kbar but remains being intermediate coupling until $40$ kbar, the highest pressure considered. In a weak coupling system the deviation function is negative definite everywhere.

\subsection{The specific heat jump at $T_c$}

It is conventional when analyzing the specific heat to add $\gamma T$, an approximation to the normal state electronic heat, onto $\Delta C$ and normalize to $\gamma T$, i.e., $C_{es}=\Delta C - \gamma T$. Table \ref{tab:table3} shows our results for the dimensionless quantity $C_{es}/\gamma T$. For this quantity the BCS weak-coupling limit at $T_c$ is 2.43. Our results with increasing pressure, are approaching, but do not reach the BCS limit (the weak coupling) as is expected from the behaviour of the calculated $D(t)$ in Fig. \ref{fig:Dv}. Our calculated values range from $2.828$ to $2.684$ for PdH at 0 and 40 Kbar respectively and from $2.928$ to $2.90$ for PdD at similar pressures.

\subsection{The gap at $T=0$}

\begin{figure}
\includegraphics[width=8.4cm]{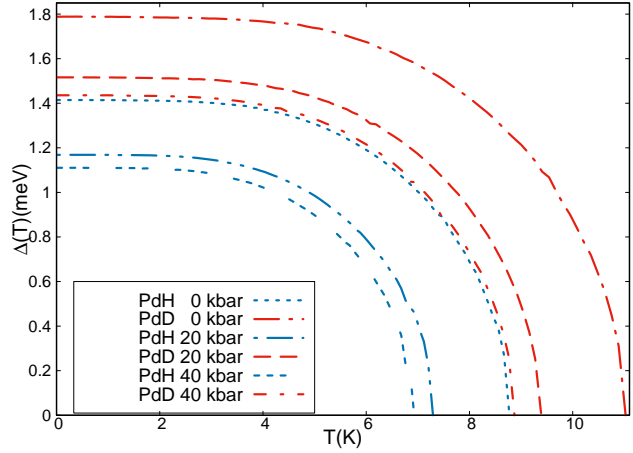}
\caption{\label{fig:gap}(color online) Superconducting gap energy.}
\end{figure}

\begin{figure}
\includegraphics[width=8.4cm]{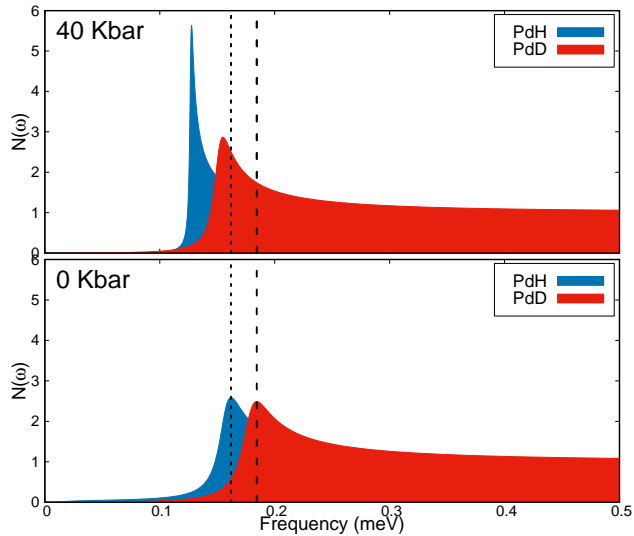}
\caption{\label{fig:Den}(color online) The quasiparticle density of states.}
\end{figure}

In order to get the quasiparticle density of states and the gap energy we perform an analytic continuation to real axis [cita]. 
The quasiparticle density of states follows from 

\begin{eqnarray}
N(\omega) = N(0)Re\left[\frac{\omega}{\sqrt{\omega^2 - \Delta^2 (\omega)}}\right]
\end{eqnarray}
Which is zero below the gap energy $\Delta_{0}$ at each temperature. Figure \ref{fig:Den} shows the quasiparticle density of states at $T_c$. For both, PdD and PdH, the energy at which the density reaches its maximum is displaced at lower energies under pressure, but the peak increases more in PdH than in PdD.
Finally, in figure \ref{fig:gap} we show $\Delta_{0}(T)$, defined by the equation 

\begin{eqnarray}
\Delta_{0}(T) = Re \left[\Delta\left(\omega=\Delta_{0}(T),T\right)\right]
\end{eqnarray}
In this case the BCS ratio $2\Delta_{0}/k_{B}T_{c}$ at $T=0$ is $3.52$. Our results ranges from $3.728$ to $3.706$ for PdH at 0 and 40 Kbar respectively and from $3.754$ to $3.747$ for PdD at similar pressures.  

\begin{table}[h]

\begin{ruledtabular}
\caption{Some thermodynamic values for PdH and PdD under pressure.}
\label{tab:table3}
\footnotesize
\begin{tabular}{ccccccc}
& \multicolumn{3}{c}{PdH} & \multicolumn{3}{c}{PdD} \tabularnewline
\hline
& 0 kBar & 20 kBar & 40 kBar & 0 kBar & 20 kBar & 40 kBar\tabularnewline
\hline
$T_c$ & 8.8 & 7.31 & 6.95 & 11.05 & 9.39 & 8.89 \tabularnewline
$2\Delta_{0}/k_{B}T_{c}$ & 3.728 & 3.708 & 3.706 & 3.754 & 3.745 & 3.747\tabularnewline
$C_{es}/\gamma T$ & 2.828 & 2.758 & 2.684 & 2.928 & 2.918 & 2.900\tabularnewline
Min. $D(t)$ & -0.0138 & -0.0160 & -0.0159 & -0.0112 & -0.01256 & -0.01252 \tabularnewline
\end{tabular}\end{ruledtabular}

\end{table}

\section{The isotope effect coefficient}

BCS theory gives for a compound with several atoms the following formula for the partial isotope effect coefficients $\alpha_{i}\equiv-d\ln T_{c}/d\ln M_{i}$ where $M_{i}$ is the mass of the different atoms in the compound and $T_{c}$ the critical temperature. The total isotope effect coefficient is given by the sum of the partial ones, namely $\alpha_{tot}=\sum_{i}\alpha_{i}$. 

A deviation from $\alpha=0.5$ could be a fingerprint of a non-conventional mechanism. For multiband superconductors this deviation is expected to vary between the BCS value and zero as long as the intraband couplings are affected. It can, however, exceed the BCS value when interband effects are dominant \cite{Bianconi8}. A multiband approach has been proposed to explain the anomalous isotope effect in pressurized $H_{2}S$ \cite{Bianconi9}. Unconventional isotope effects can also be due to lattice fluctuations. This fluctuations have been identified in cuprates as due to the role of the lattice in the pseudogap formation \cite{Bianconi1,Bianconi2} or to the modification of the electron-phonon interaction in systems such as $Mg_{1-x}Al_{x}B$ \cite{Bianconi7} and $YB_{2}$ \cite{Bianconi6}. It has also been shown that systems with anomalous isotope coefficients have anomalies at Lifshitz transitions \cite{Bianconi3,Bianconi4,Bianconi5}. The effects of the zero point motion has been proposed to explain the anomalous isotope effect in $H_{3}S$ \cite{Bianconi10}.

In palladium hydride the replacement of hydrogen by deuterium results in a higher superconducting temperature and in an anomalous isotope effect ($\alpha\approx-0.3$). For this system, most of the mechanism that have been proposed to explain it attribute the inverse isotope effect to vibrational effects of Hydrogen and Deuterium, such as anharmonicity \cite{Kara,PhysRevB.45.12405, PhysRevLett.111.177002} or to the zero-point motion \cite{PhysRevB.29.4140}, both of which have an effect on the electron-phonon coupling without reaching to a satisfactory explanation of this phenomenon. For example, Errea et. al. \cite{PhysRevLett.111.177002} study the anharmonicity of the hydrogen vibrations within the rocksalt crystal structure and found the value of the isotope coefficient to be $\alpha=-0.38$, however their calculated critical temperatures does not match the experimental ones (see Tab. \ref{tab:table2}). In the work of Jena et. al. \cite{PhysRevB.29.4140} they show that the electronic structure of PdH and of PdD is influenced by the zero-point vibration of hydrogen and deuterium. Even more, Yussouff et. al. \cite{YUSSOUFF1995549} included anharmonicity on top of the zero-point effects and found the isotope coefficient to be $\alpha=-0.3926$. 

More recently, S. Villa-Cort\'es and R. Baquero [auto cita] have been proposed a different approach to explain the inverse isotope effect in palladium hydride. They considered the vibrational modes to be harmonnic and the crystal structure to be zincblende. In their approach they split the isotope coefficient into two contributions. One that comes from the change in the electron-phonon interaction, namely $\alpha^{el-ph}$, and other that comes from the change in the electron-electron interaction, namely $\alpha^{el-el}$. The last one gives us information on how the screening of the electron-electron interaction is affected by the isotope substitution. At 0 pressure they found the electron-phonon contribution to be $\alpha^{el-ph}=0.0556$ and the electron-electron contribution to be $\alpha^{el-el}=-0.369$, this gives a total isotope coefficient $\alpha= -0.3134$ in remarkable agreement to the experimental value (see Tab. \ref{tab:table2}). For a detailed account of this approach see [autocita]. Here we extend our previous work to include the effects of pressure in the isotope effect.

\begin{figure}
\includegraphics[width=8.4cm]{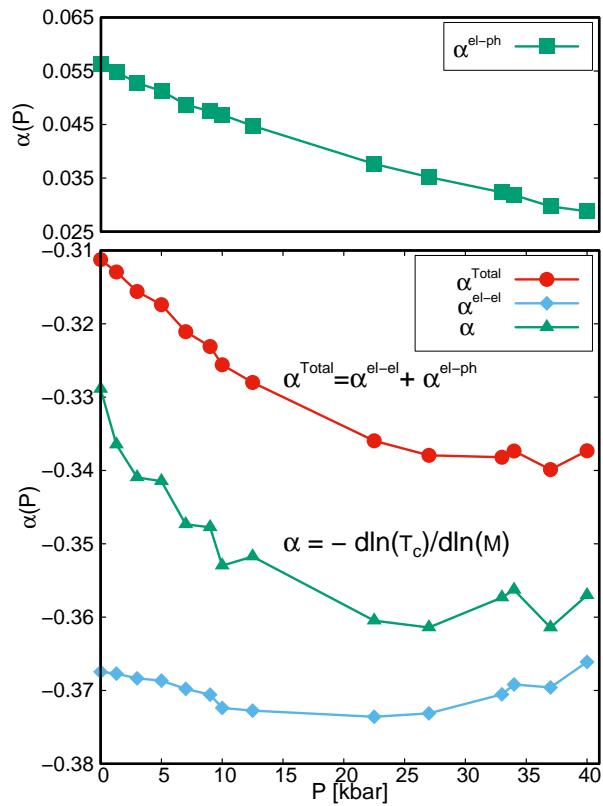}
\caption{\label{fig:alfa}(color online) Calculated isotope coefficient for PdH(D) under pressure. The $\alpha$ values (triangles) were calculated using the $T_c$ under pressure from Fig. \ref{fig:Tc-P}.}
\end{figure}

In Fig. \ref{fig:alfa} we show our results for the isotope coefficient under pressure. We procced in two ways. First we use the calculated $T_c$ under pressure for PdH and PdD (see Fig. \ref{fig:Tc-P}) to calculate the isotope coefficient from $\alpha(P)=-d\ln T_{c}(P)/d\ln M$. In general it tends to diminish under pressure. Second, we calculated both contributions for the isotope effect, the electron-electron and the electron-phonon interactions. In general $\alpha^{Total}$ follows closely the same trend of the calculated isotope effect under pressure. In general the magnitude of the inverse isotope coefficient can be explained by the change in the electron-electron interaction, $\alpha^{el-el}$, but the slight variation under pressure comes from the variation of the electron-phonon interaction under pressure.

\begin{table}
\begin{ruledtabular}

\caption{Superconducting critical temperature and isotope effect coefficient from experimental and calculated results.}
\label{tab:table2}
\footnotesize
\begin{tabular}{cccc}
& \multicolumn{2}{c}{$T_c$ K} & $\alpha$ \tabularnewline
\hline
& PdH & PdD & \tabularnewline
\hline
This work & 8.82 & 11.05 &  -0.3134 \tabularnewline
Expt. \cite{PhysRevB.39.4110} & 8.8 & 11.05 &  -0.3289 \tabularnewline
Expt. \cite{PhysRevB.10.3818} & 8 & 10 & -0.3222\tabularnewline
Expt. \cite{PhysRevLett.40.578} & 9.5 & 11.6 & -0.2883\tabularnewline
From \cite{PhysRevLett.111.177002} & 5 & 6.5 & -0.38\tabularnewline
From \cite{YUSSOUFF1995549} & 8 & 10.5 & -0.3926\tabularnewline

\end{tabular}\end{ruledtabular}
\end{table}
 
\section{CONCLUDING REMARKS}

In conclusion, we have used a novel method \cite{ivan} to calculate the superconducting thermodynamics under pressure within the Migdal-Eliashberg theory. We have found that if we consider the zincblende crystal structure as the proper one for PdH and PdD at low temperatures ($T\ll55$ K) in agreement with experiment, the experimental values of the critical temperature and of the inverse isotope effect under pressure are well reproduced. This method represents a basis on which the thermodynamics of other hydrides under pressure can be calculated. We have found that the enhancement of the critical temperature, and then the inverse isotope effect, can be explained by taking the screening of the electron-electron interaction by the phonons into account. Further we have calculated the dependence under pressure of the specific heat at $T_c$, the energy gap at $T=0$, the deviation function $D(t)$ and the density of the quasiparticle states. As a general remark we can say that the changes induced by pressure in the thermodynamics tends to weaken the system.
    
\begin{acknowledgments}
The authors acknowledge to the general coordination of information and communications technologies (CGSTIC) at CINVESTAV-IPN for providing HPC resources on the Hybrid Cluster Supercomputer "Xiuhcoatl", that have contributed to the research results reported within this paper. S. Villa-Cort\'es acknowledges the support of Conacyt-M\'exico through a PhD scholarship.
\end{acknowledgments}

\section*{References}
\bibliographystyle{unsrt}
\bibliography{PEII_IOP}

\end{document}